\pdfoutput=1
\documentclass{JINST}
\usepackage{amsmath}
\usepackage[utf8]{inputenc}
\usepackage{url}
\newcommand{\unit}[1]{\ensuremath{\, \mathrm{#1}}}

\title{Primary Beam Steering Due to Field Leakage from Superconducting SHMS Magnets}

\author{Michael H. Moore$^{ab}$\thanks{Corresponding author.}~ ,
Buddhini P. Waidyawansa$^a$, Silviu Covrig$^a$, Roger Carlini$^a$
and Jay Benesch$^a$\\
\llap{$^a$}Thomas Jefferson National Accelerator Facility \\Newport News, VA 23606 USA\\
\llap{$^b$}Old Dominion University\\ Norfolk, VA 23508 USA\\ 
E-mail: \email{mhmoore@jlab.gov}}

\abstract{
Simulations of the magnetic fields from the Super High Momentum Spectrometer in Hall C at Thomas Jefferson National Accelerator Facility show significant field leakage into the region of the primary beam line between the target and the beam dump. Without mitigation, these remnant fields will steer the unscattered beam enough to limit beam operations at small scattering angles. Presented here are magnetic field simulations of the spectrometer magnets and a solution using optimal placement of a minimal amount of shielding iron around the beam line.
} 

\keywords{
Spectrometers; Detector modeling and simulations II; Accelerator modelling and simulations
} 
\begin{document}

\section{Introduction}

The Super High Momentum Spectrometer (SHMS) was designed for operations up to 11 GeV in Hall C at Thomas Jefferson National Accelerator Facility (JLab)~\cite{hallc}. In Hall C an electron beam is incident upon a fixed target and scattered particles are analyzed by two magnetic spectrometers. The remaining unscattered beam then travels past the spectrometer to a high power beam dump ($51.8$ m from the target) where it can be safely absorbed~\cite{sinclair}. Recent simulations have shown that the magnetic field leakage from the SHMS magnets would steer the unscattered primary beam away from the center of the beam dump~\cite{carlini}. The acceptable region of the beam dump for high current operation is a relatively small area in the center of the larger water tank which comprises the beam dump. If unmitigated, SHMS operations would be restricted to low electron-beam currents for small spectrometer angles. Presented here are the first 3D magnetic field simulations of the four leading magnets of the SHMS with emphasis placed on the fields along the beam line and the position of the beam at the beam dump due to these fields. A passive solution using optimal placement of extra iron along the beam line is presented.

\section{Geometry of the SHMS}
\begin{figure}[!htbp]
	\centering
		\includegraphics[width=0.8\textwidth]{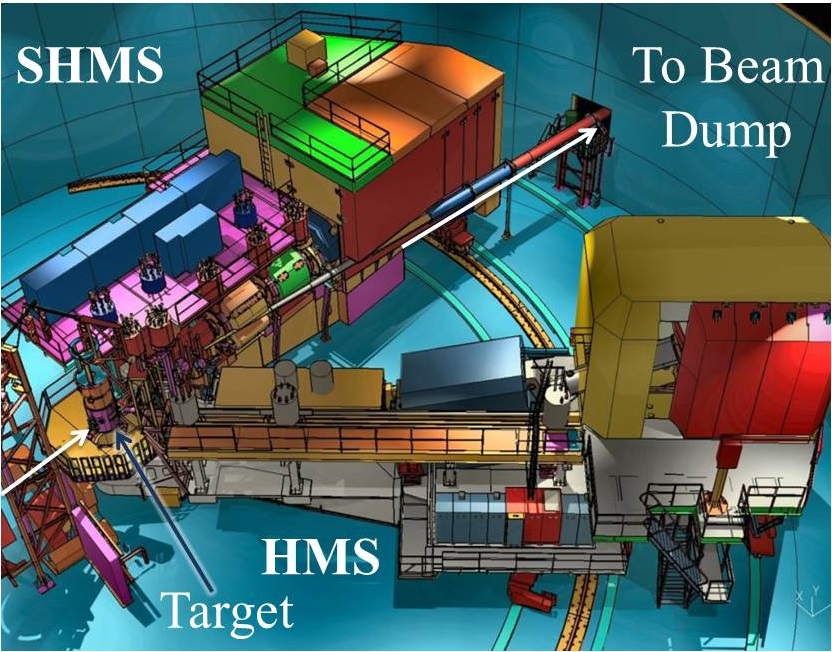} 
		\caption{Hall C at JLab showing the two spectrometers and the primary beam line. The primary beam line is indicated by white arrows.}
	\label{fig:hallc}
\end{figure}

\begin{figure}[!htbp]
	\centering
		\includegraphics [width=0.8\textwidth] {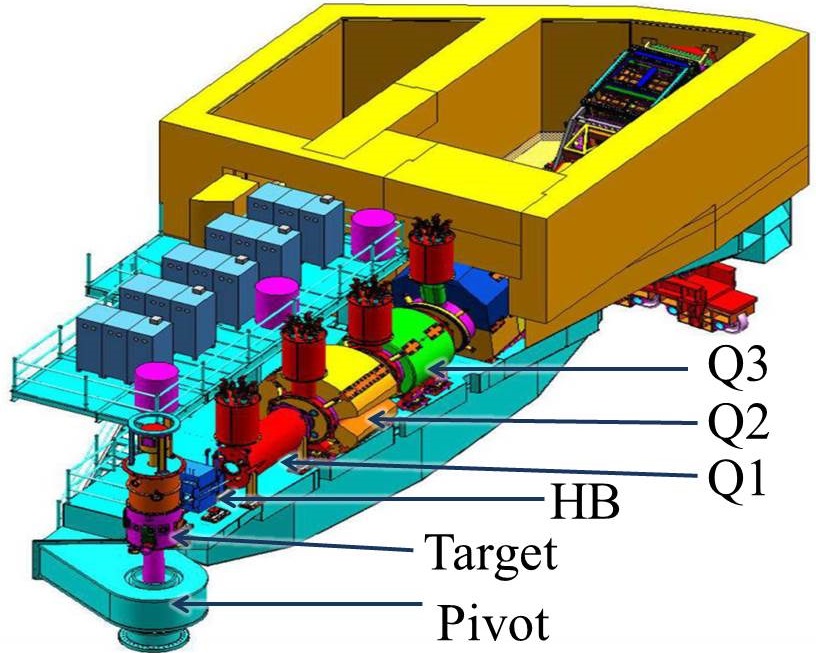} 
		\caption{Main elements of the SHMS. See text for explanation.}
	\label{fig:shms}
\end{figure}

The existing High Momentum Spectrometer (HMS) will be used jointly with the new SHMS. Both the HMS and the new SHMS are connected to a central pivot located underneath the target chamber (see Figure~\ref{fig:hallc}). Large wheels rolling on rails around the target allow these spectrometers to accommodate a wide range of scattering angles. The HMS has an angular range of $12.5^{\circ}$ to $~90^{\circ}$ and the SHMS has a range of $5.5^{\circ}$ to $40^{\circ}$.  One of the primary differences between the HMS and the newly designed SHMS is a $3^{\circ}$ horizontal bending magnet (HB) placed between the target and the first quadrupole magnet. When positioned at the smallest scattering angle, the SHMS makes an angle of $8.5^{\circ}$ with respect to the beam line. The HB allows particles scattered at an angle of $5.5^{\circ}$ to be steered into the optics of the SHMS. Both spectrometers have focusing elements consisting of quadrupole triplets (Q1, Q2, Q3)~\cite{brindza}. Figure~\ref{fig:shms} shows the elements of the SHMS used in this analysis.


\begin{figure}[!htbp]
	\centering
		\includegraphics [width=1.0\textwidth] {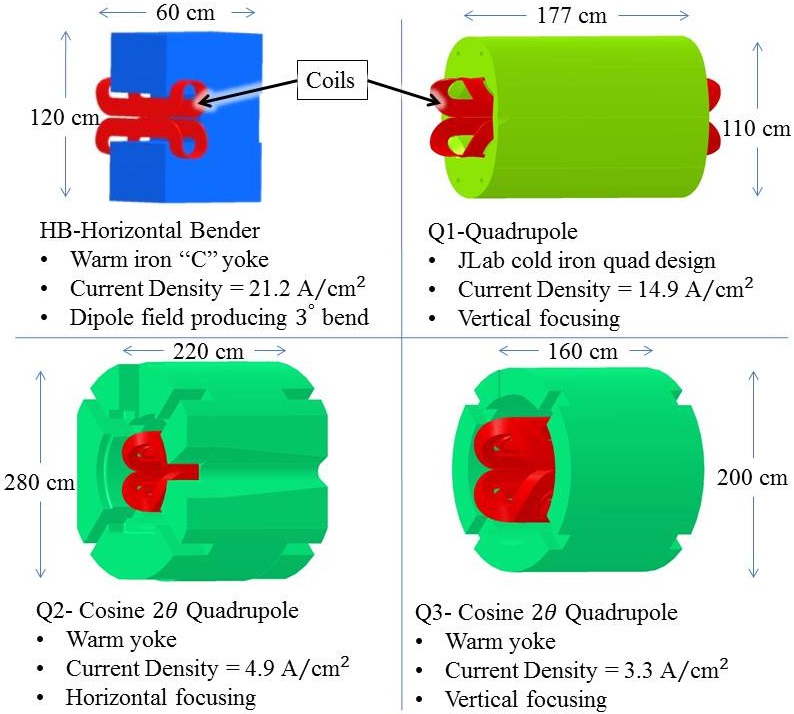} 
		\caption{SHMS magnets as built in Opera. The current densities are for the maximum central momentum setting of the spectrometer.}
	\label{fig:mags}
\end{figure}

\begin{figure}[!htbp]
	\centering
		\includegraphics [width=1.0\textwidth] {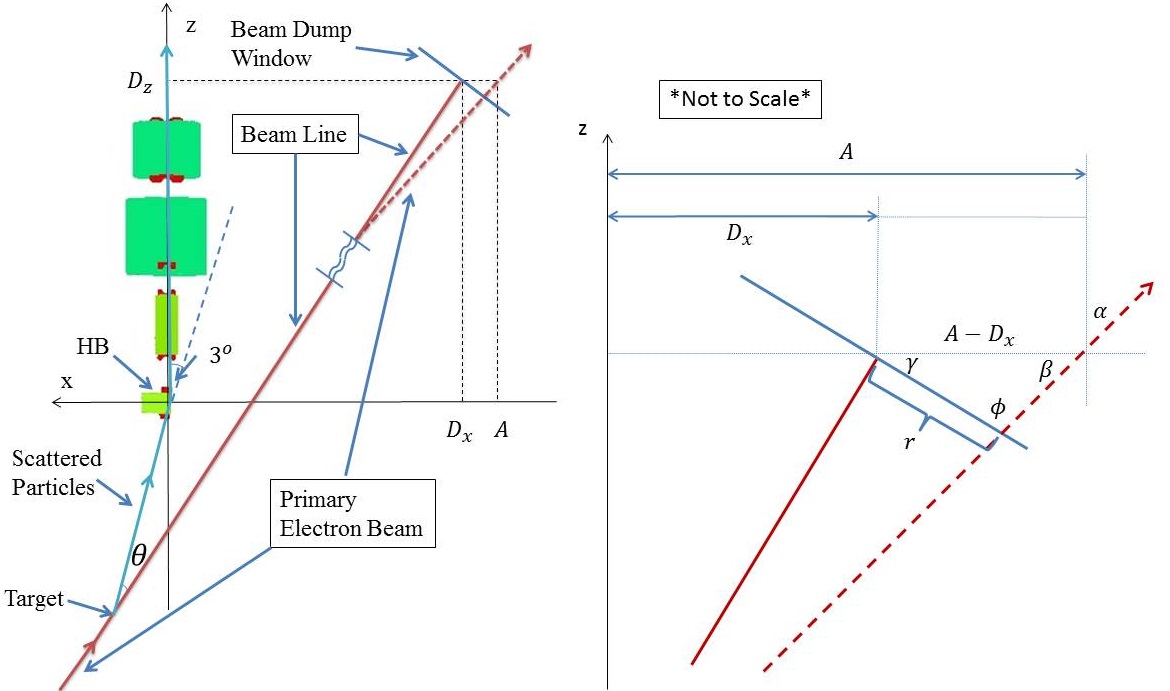} 
		\caption{\emph{Left}:The geometry of the beam line on the coordinate system with the origin at the HB and the $z$-axis along the optics line. The primary beam trajectory follows the beam line to the beam dump window with no field leakage (solid line). With field leakage (dashed line) it misses the center.  \emph{Right}: Close up of the intersection between the beam line and the beam dump window showing the beam displacement, $r$. See text for explanation.}
	\label{fig:geo}
\end{figure}

In order to simulate the magnetic fields from the SHMS,  the yokes and coils of the magnets were modeled with the Opera-3D software~\cite{opera} as shown in Figure \ref{fig:mags}. Opera serves as the modeler and post processor for a suite of finite element simulation tools. TOSCA is the static electromagnetic field simulator used with Opera to find the magnetic fields in this study. The coordinate system of the simulation was chosen with the $z$-axis pointing downstream along the optics line of the SHMS (through the center of the quads) and the origin placed at the vertex of the $3^{\circ}$ bend in the HB (see the left diagram in Figure~\ref{fig:geo}).

The spectrometers can be set to accept either positive or negative particles by reversing the polarity of the magnets. 
TOSCA simulations of fields for positive and negative spectrometer settings result in equal and opposite magnetic fields along the beam line. To avoid complication only negative spectrometer settings are reported.

Once the fields are found using TOSCA, to determine the beam displacement at the dump, a single electron trajectory was calculated using Opera's post-processor. Given the electron's initial position (the target), beam energy (11\,GeV), and direction (along the beam line), Opera calculates the Lorentz force in 1\,cm steps along the trajectory and returns the electron's position, velocity and time at every step. Using this information and the geometry from Figure~\ref{fig:geo}, the beam displacement at the beam dump can be estimated as follows. 

In the simulation coordinate system the target and the center of the beam dump window are located in the $(x,z)$ plane. With the center of the target fixed at $(x=9.21, z=-175.76) \unit{cm}$, the center of the beam dump window is found as a function of the scattering angle, $\theta$,

$$D_x(\theta)=-(D\sin{(\theta +3)}-9.21),$$ 
$$D_z(\theta)=D\cos{(\theta +3)}-175.76,$$ 

where $D=5180$ cm is the distance from the target to the center of the beam dump window. Since the beam dump window is always perpendicular to the beam line,  the angle of the window also changes with $\theta$ in this model. The diagram on the right side of Figure~\ref{fig:geo} shows the geometry used for this correction. The law of sines for the triangle in the diagram states 

$$\frac{A-D_x}{\sin{\phi}}=\frac{r}{\sin{\beta}}. $$

Then using the relations $\beta=90-\alpha$ and $\phi=90-(\gamma-\alpha)$, the displacement $r$ at the dump is
 
\begin{equation}
r(A, \alpha)=(A-D_x)\frac{\cos{(\alpha)}}{\cos{(\gamma-\alpha)}},\label{eq:r}
\end{equation}

where $D_x$ and $\gamma$ are determined from the spectrometer angle. By assuming that the magnetic fields at the beam dump window are negligible, it is possible to calculate $A$ and $\alpha$ from the Opera trajectory file. Note that a negative displacement means beam right and positive is beam left.

For the maximum displacement ($r_{\unit{max}}$) of a single electron at the beam dump window, allowances must be made for the growth in the size of the beam due to square rastering before the target and scattering at the beam diffuser and several vacuum windows. For safe operations of the beam dump, the radiation control group at Jefferson Lab suggests the beam size at the dump window should be set to a $4\times4 \unit{cm}^2$ square around the center of the beam~\cite{jay}. This limits $\left|r_{\unit{max}}\right|$ to $2.25$ cm so that the $4\times4 \unit{cm}^2$ rastered beam profile fits within the $5.08$ cm radius high current acceptance region of the beam dump window.

\section{Results For SHMS As Designed}

\begin{figure}[!htbp]
	\centering
		\includegraphics [scale=0.5] {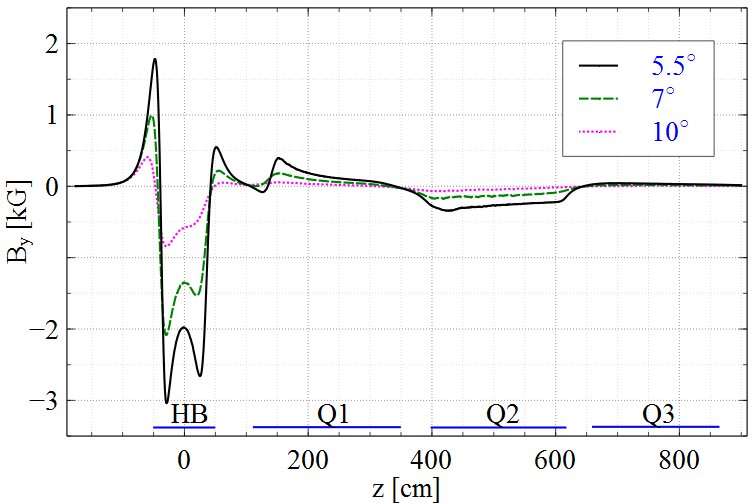} 
		\caption{Variation of the vertical component, $B_y$, of the simulated field along the beam line with the as designed SHMS. Three spectrometer angles are shown with the spectrometer central momentum set to a beam energy of 11~Gev. The horizontal axis gives the $z$-component of the distance from the target along the beam pipe. The edges of the four magnet yokes are marked along the horizontal axis.}
	\label{fig:By}
\end{figure}

From the fields simulated by TOSCA, it was immediately evident that the only field component of interest was the vertical component along the beam line. Figure~\ref{fig:By} shows the variation of the vertical component of the field along the beam line for several spectrometer angles at 11\,GeV. For a spectrometer angle of $5.5^{\circ}$ at $11$ GeV the magnetic field integral, $\int B_y dl$, of the HB field leakage along the beam line ($z=\{-100,100\}\unit{cm}$) is responsible for $78\%$ of the total field leakage.
 
\begin{figure}[!htbp]
	\centering
		\includegraphics [scale=0.5] {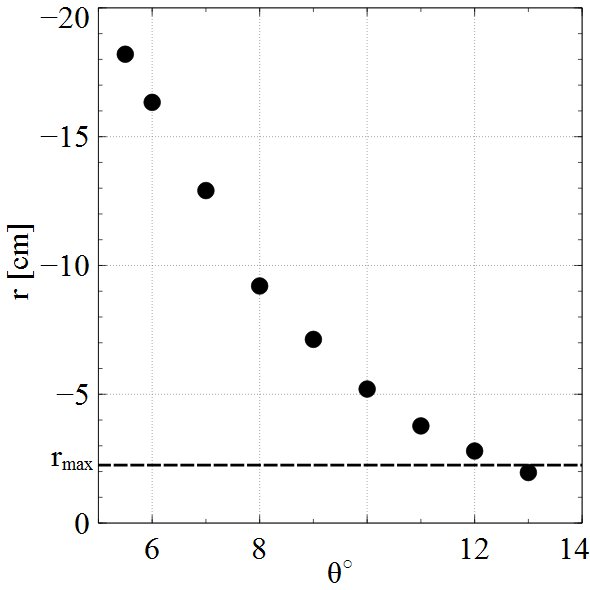} 
		\caption{Beam displacement, $r$, from the center of the beam dump window vs spectrometer angle. All displacements are taken with a beam energy of $E=11$ GeV and the spectrometer central momentum set to the beam momentum. The dashed line indicates the maximum allowed displacement, $r_{\unit{max}}=2.25$\,cm. Everything above this line will miss the acceptable region of the beam dump window. Negative $r$ values correspond to beam right displacements.}
	\label{fig:A_displacements}
\end{figure}

\begin{figure}[!htbp]
	\centering
		\includegraphics [scale=0.5] {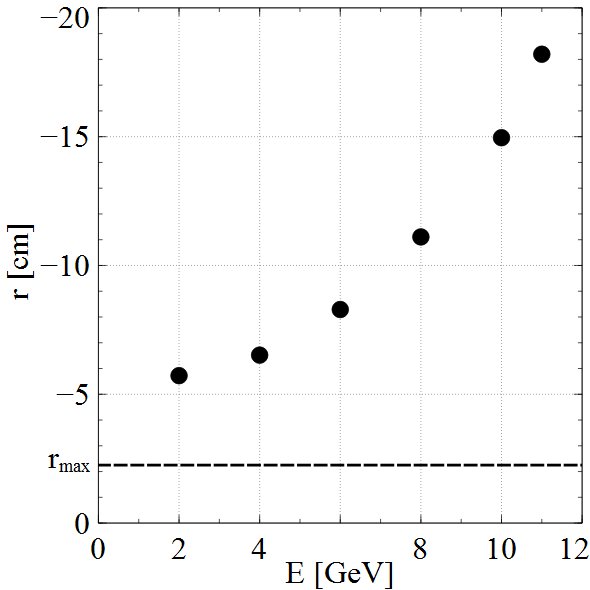} 
		\caption{Beam displacement, $r$, from the center of the beam dump window vs beam energy with the spectrometer central momentum set to the beam momentum. The displacements are taken with $\theta=5.5^{\circ}$. The dashed line indicates the maximum allowed displacement, $r_{\unit{max}}= 2.25$\,cm. Everything above this line will miss the acceptable region of the beam dump window. Negative $r$ values correspond to beam right displacements.}
	\label{fig:E_displacements}
\end{figure}

The displacement of the beam calculated from Eq.(\ref{eq:r}) for the angles and energies of interest are plotted in Figures~\ref{fig:A_displacements} and~\ref{fig:E_displacements} . With the as-designed SHMS, the primary beam has a displacement at the beam dump window of $-18.2$\,cm when operated at $5.5^{\circ}$ and $11$\,GeV, and misses the window at all angles below $13^{\circ}$. The displacement vs beam energy in Figure~\ref{fig:E_displacements} has the central momentum of the spectrometer set to match the incident beam energy. The displacements were more than $r_{\unit{max}}$ for all the energies studied, $E=\{2,4,6,8,10,11\}$ GeV. These displacements would inhibit beam operations in Hall C, therefore a solution needed to be found that would bring the displacements below $|r_{\unit{max}}|=2.25$ cm for all angles and energies.

\section{Passive Solution}
Redesigning of the SHMS magnets to overcome the beam steering issue is not practical due to schedule and cost constraints. At the beginning of this analysis the magnets were already being manufactured. The first beam in Hall C is scheduled for 2016.

Two possible solutions remain: (1) Use steering magnets, accompanied by beam position monitors (BPMs), to actively steer the beam back to the center of the beam dump window or (2) add extra iron to the existing magnet yokes and primary beam pipe to reduce the field leakage along the beam line as a passive solution.

Steering magnets and BPMs located downstream of the target would have to be radiation hardened and operate continuously making this the most expensive solution. A steering magnet could be placed before the target but this would affect the optics. Any active solution would have to be re-tuned with every change of scattering angle or energy. The passive solution has several advantages over steering magnets, most importantly it is intrinsically fail-safe. The fields in the beam pipe region will be attenuated regardless of spectrometer setting. However, extra material downstream of the target presents an activation hazard and steps must be taken to minimize activation. These steps include minimizing the amount of material and modeling the activation using GEANT4~\cite{buddhini}. 

\begin{figure}[!htbp]
	\centering
		\includegraphics [scale=.4] {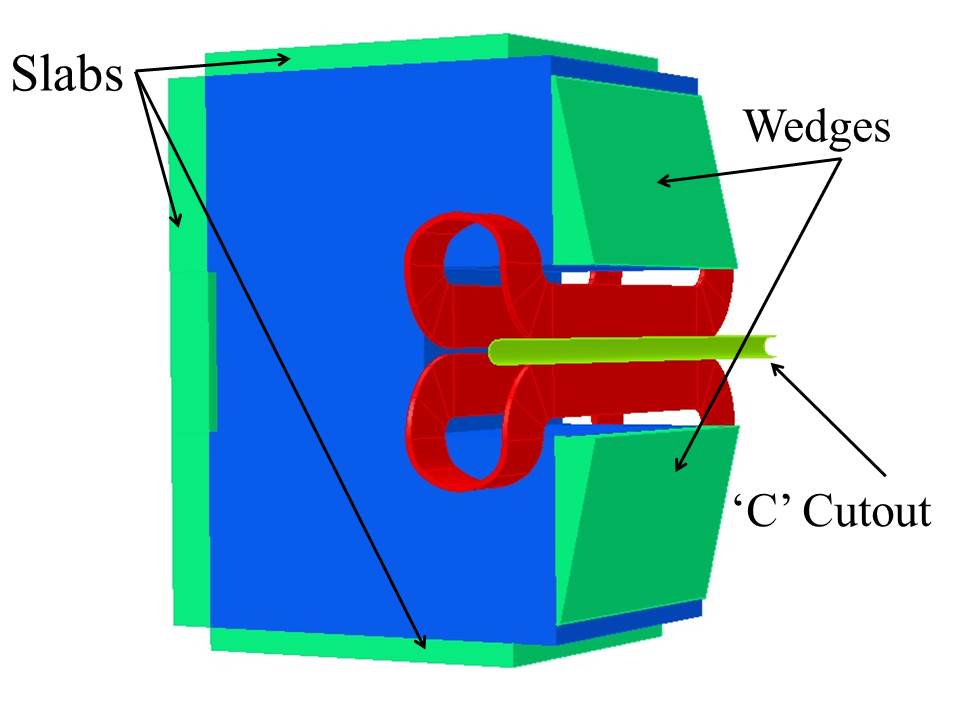} 
		\caption{Extra iron pieces added to the HB yoke to reduce field leakage. The slabs and wedges are attached directly to the yoke. The 'C' cutout is permanently attached to the cryostat which houses the coils.}
	\label{fig:extrairon}
\end{figure}

As a first step in simulating the fields for the passive solution, extra iron of various shapes and sizes were added to the HB yoke. The most advantageous was clearly an iron pipe around the beam pipe but such a large amount of material so close to the beam line was undesirable due to activation. The iron additions that were most successful are shown in Figure~\ref{fig:extrairon} and the corresponding reduction in field integrals are given in Table~\ref{table:extrairon}. The field integrals are along the beam line from $z=\{-176, 900\}$ with a $5.5^{\circ}$ scattering angle and the maximum central momentum setting of the spectrometer. The wedges are the most successful at attenuating the fields and are far enough away from the beam line to not become an activation problem.

\begin{table}[ht]
	\caption{Field integral reductions from adding extra iron to the HB yoke.} 
	\centering 
		\begin{tabular}{c c c  } 
	\hline
		Additions & Field Integral (kG cm)& $\%$ Reduction\\
	\hline 
		None		& -110.49		&0		\\
		Slabs	&-108.76		&1.57	\\
		Wedges	&-58.82		&46.77	\\
		'c' cutout	&-89.20		&19.27	\\
	\hline
		All		&-35.80		&67.6	\\

		\end{tabular}

	\label{table:extrairon} 
\end{table}

\begin{figure}
	\centering
		\includegraphics [scale=0.5] {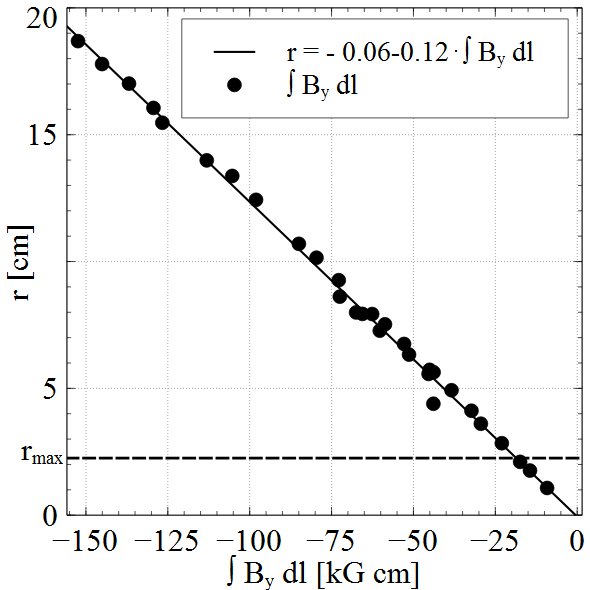} 
		\caption{Beam displacement, $r$, vs field integral for 30 different combinations of extra iron pieces on the HB yoke and pipes along the beam line. The linear fit determines the maximum field integral for a displacement, $r<r_{\unit{max}}$. All the data are taken for a $5.5^{\circ}$ scattering angle, 11\,GeV electron beam and maximum central momentum setting of the spectrometer.}
	\label{fig:DvsBdl}
\end{figure}

For comparison, the displacements of 30 different combinations of extra iron pieces and pipes are plotted versus their magnetic field integrals in Figure~\ref{fig:DvsBdl}. All these simulations are for a $5.5^{\circ}$ scattering angle, 11\,GeV electron beam and maximum central momentum setting of the spectrometer. The linear fit determines the maximum absolute value of the field integral that constrains the displacement at the the beam dump to $r_{\unit{max}}=\pm 2.25$ cm. Under these conditions, the absolute value of the field integral must be less than $19.25 \unit{kG\cdot cm}$.

Unfortunately, all the extra iron pieces on the HB yoke are unable to constrain the absolute value of the field integral to below $19.25$ kG$\cdot$cm. To further reduce the field integral, an iron pipe around the beam line at the HB was then studied as the next step of the simulation.

\begin{figure}
	\centering
		\includegraphics [width =1.0\textwidth] {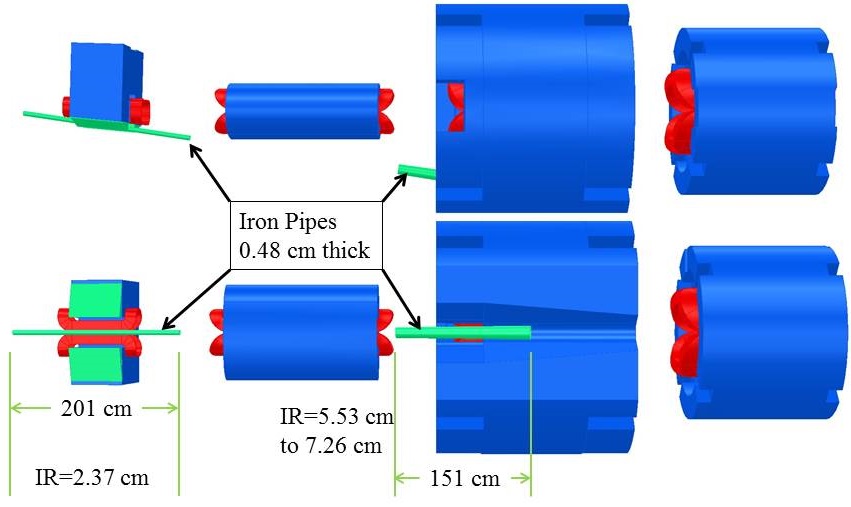} 
		\caption{Top \emph{(top)} and side view (\emph{bottom}) of the first four SHMS magnets showing the iron wedges on the HB yoke and iron pipes around the beam line at HB and Q2.}
	\label{fig:pipe}
\end{figure}

Several thicknesses, lengths and positions of pipes were studied in Opera. It was found that when used in conjunction with the wedges on the HB, the thinnest pipe wall that provides effective shielding of the field was $0.476$ cm (3/16"). Initially, short pipes around the HB were studied to balance the total field integral. Although the displacement for $5.5^{\circ}$ could be minimized to below $r_{\unit{max}}$, the behavior at larger angles and smaller energies was erratic mainly due to saturation effects and changing locations of field maximums at different angles. To make the pipe solution robust at all angles and energies, a 2\,m long pipe at HB and a 1.5\,m pipe at Q2 were designed (see Figure~\ref{fig:pipe}). The HB pipe has an inner radius (IR) of 2.37\,cm through the entire length. To minimize activation, the Q2 pipe is tapered from IR=5.53\,cm on the upstream end to IR=7.46\,cm on the downstream end.

\begin{figure}
	\centering
		\includegraphics [scale=0.5] {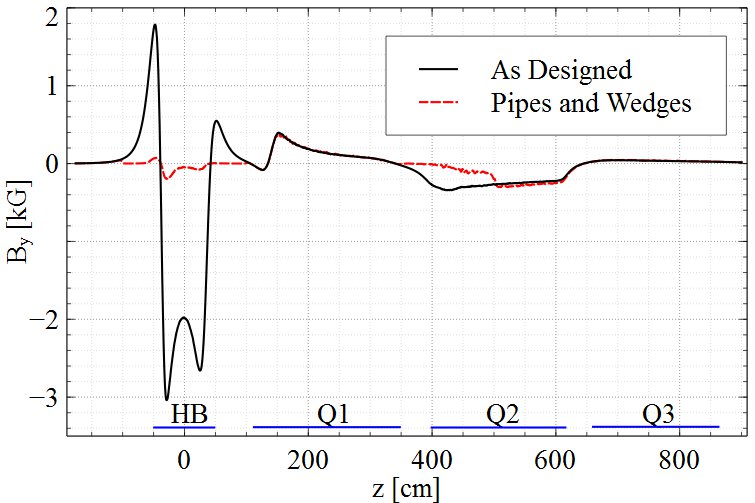} 
		\caption{Vertical component $B_y$ of the simulated field along the beam line from the as-designed SHMS and the SHMS with pipes and wedges at $\theta=5.5^{\circ}$ and $11$\,GeV.}
	\label{fig:Bypipe}
\end{figure}

\begin{figure}
	\centering
		\includegraphics [scale=0.5] {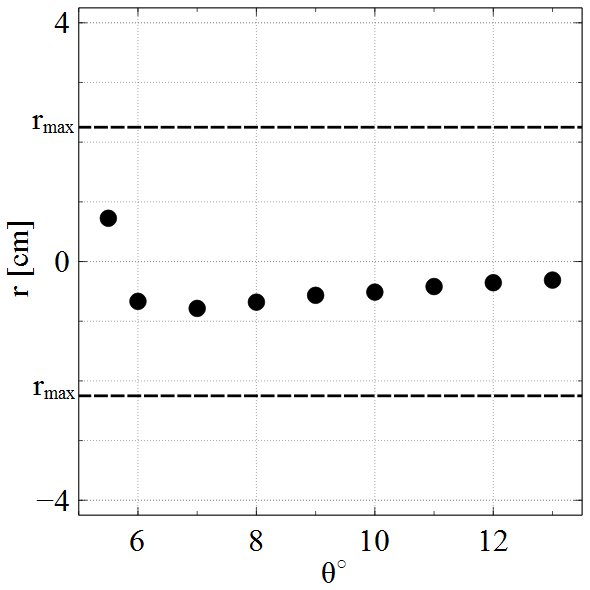} 
		\caption{Beam displacement, $r$, from the center of the beam dump window vs spectrometer angle when using iron wedges and pipes. All displacements are taken with a beam energy of $E=11$ GeV and the spectrometer central momentum set to the beam momentum. The dashed lines indicates the maximum allowed displacements, $r_{\unit{max}}=\pm 2.25$\,cm. Everything outside these lines will miss the acceptable region of the beam dump window. Positive $r$ values correspond to beam left displacements. This solution works at all angles studied.} 
	\label{fig:A_pipedisplace}
\end{figure}

\begin{figure}[!htbp]
	\centering
		\includegraphics [scale=0.5] {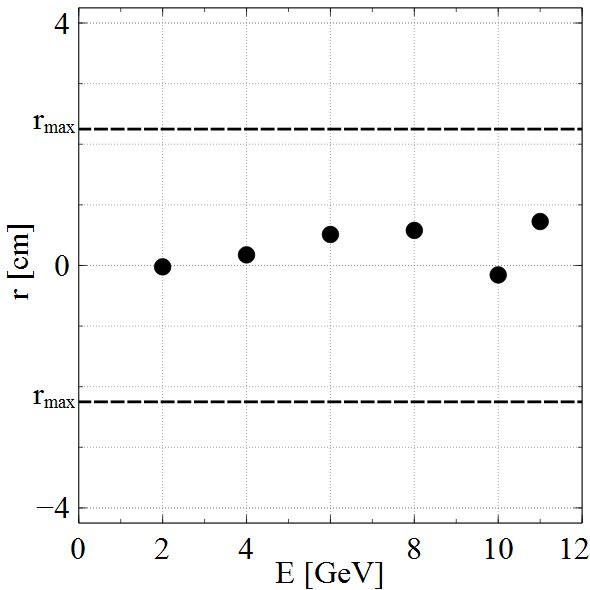} 
		\caption{Beam displacement, $r$, from the center of the beam dump window vs beam energy with the spectrometer central momentum set to the beam momentum. The displacements are taken with $\theta=5.5^{\circ}$. The dashed lines indicate the maximum allowed displacements, $r_{\unit{max}}=\pm 2.25$\,cm. Everything outside these lines will miss the acceptable region of the beam dump window. Positive $r$ values correspond to beam left displacements. This solution works at all energies studied.} 
	\label{fig:E_pipedisplace}
\end{figure}

The simulated magnetic field along the beam line with the iron wedges and pipes  is shown in Figure~\ref{fig:Bypipe}. These simulations are for a $5.5^{\circ}$ scattering angle and maximum central momentum setting of the spectrometer. The as-designed SHMS field is shown for comparison. The magnetic field integral along the beam line with the iron pipes and wedges is $-6.8$ kG$\cdot$cm, meeting the criteria of an absolute value less than $19.25$ kG$\cdot$cm found above. The high attenuation of the entire HB external field makes this solution robust and fail-safe. If there is any problem with the magnets (power supply trip-off, quench, etc.), the primary beam will still have a displacement at the beam dump window which is less than $r_{\unit{max}}$. The beam displacements at the beam dump window from this solution are less than $|r_{\unit{max}}|=2.25$ cm at all energies and angles (see Figures~\ref{fig:A_pipedisplace} and ~\ref{fig:E_pipedisplace}).
 
\section{Summary}
The Opera models and TOSCA solutions presented here have shown themselves to be valuable for simulating the fields of the optics system in the SHMS. 

The SHMS, as it is being built, has an operational minimum scattering angle of $13^{\circ}$ at $11$ GeV due to field leakage from the first four magnets. This is $7.5^{\circ}$ more than the designed minimum scattering angle of $5.5^{\circ}$. The stray fields steer the unscattered primary beam by as much as $18.2$ cm from the center of the beam dump window. This is $15.95$ cm farther than the suggested maximum for safe beam operations in Hall C.   

The passive solution consisting of pipes and wedges suggested here is a robust and fail-safe solution that works at all angles and energies. For scattering angles of $5.5^{\circ}$ to $10^{\circ}$ and a beam energy of $11$\,GeV, the beam always hits the beam dump window within $1$ cm of the center. The same can be said for beam energies with matching spectrometer central momentum from $2$ to $11$ GeV when the spectrometer is operated at a $5.5^{\circ}$ scattering angle. At larger spectrometer angles (greater than $13^{\circ}$) the iron pipes can be removed. Using clam-shell style sheaths instead of pipes would allow for easy removal and replacement. 

\section*{Acknowledgments}
Notice: This manuscript has been authored by Jefferson Science Associates, LLC under Contract No. DE-AC05-06OR23177 with the U.S. Department of Energy. The United States Government retains and the publisher, by accepting the article for publication, acknowledges that the United States Government retains a non- exclusive, paid-up, irrevocable, world-wide license to publish or reproduce the published form of this manuscript, or allow others to do so, for United States Government purposes. The authors wish to thank Howard Fenker, Bert Meztger and Stephen Wood for their outstanding support.

\bibliography{Master}{}
\bibliographystyle{JHEP}

\end{document}